\documentclass[a4paper]{jpconf}

\begin{document}

\title{Emergent gravity in graphene
\\{\small \it dedicated to the memory of M.I.Polikarpov}
}

\author{M.A.~Zubkov}

\address{ITEP, B.Cheremushkinskaya 25, Moscow, 117259, Russia
}

\author{G.E.~Volovik}

\address{Low Temperature Laboratory, School of Science and
Technology, Aalto University,  P.O. Box 15100, FI-00076 AALTO, Finland;

L. D. Landau Institute for Theoretical Physics,
Kosygina 2, 119334 Moscow, Russia}

%\ead{zubkov@itep.ru}

\begin{abstract}
We reconsider monolayer graphene in the presence of elastic deformations. It is described by the
tight - binding model with varying hopping parameters. We demonstrate, that the fermionic quasiparticles propagate in the
emergent $2D$ Weitzenbock geometry and in the presence of the emergent $U(1)$ gauge field. Both emergent geometry and the gauge field are defined by the elastic deformation of graphene.
\end{abstract}

%%%%%%%%%%%%%%%%%%%%%%%%%%%%%%%%%%%%%%%%%%%%%%%%%%%%%%%%%%%%%%%%%%%%%%%%

%%%%%%%%%%%%%%%%%%%%%%%%%%%%%%%%%%%%%%%%%%%%%%%%%%%%%%%%%%%%%%%%%%%%%%%

%\maketitle

%\documentclass[a4paper,11pt]{article}
%\pdfoutput=1 % if your are submitting a pdflatex (i.e. if you have
             % images in pdf, png or jpg format)

%\usepackage{jheppub} % for details on the use of the package, please
                     % see the JHEP-author-manual

%\usepackage[T1]{fontenc} % if needed

%\usepackage{slashed}
%\usepackage[dvips]{color}
%\usepackage[dvips]{epsfig}
%\usepackage{latexsym}
%\usepackage{bm}
%\usepackage{upgreek}
%\usepackage{mathrsfs}
%\usepackage{times}
%\usepackage{amsthm}
%\usepackage{amssymb}
%\usepackage{epsfig}
%\usepackage{graphicx}
%\usepackage{amsmath}

\newcommand{\barray}{\begin{eqnarray}}
\newcommand{\earray}{\end{eqnarray}}
\newcommand{\nn}{\nonumber \\}
\newcommand{\nl}{& \nonumber \\ &}
\newcommand{\bnl}{\right .  \nonumber \\  \left .}
\newcommand{\dbnl}{\right .\right . & \nonumber \\ & \left .\left .}

%Begin-end
\newcommand{\beq}{\begin{equation}}
\newcommand{\eeq}{\end{equation}}
\newcommand{\ba}{\begin{array}}
\newcommand{\ea}{\end{array}}
\newcommand{\bea}{\begin{eqnarray}}
\newcommand{\eea}{\end{eqnarray} }
\newcommand{\be}{\begin{eqnarray}}
\newcommand{\ee}{\end{eqnarray} }
\newcommand{\bal}{\begin{align}}
\newcommand{\eal}{\end{align}}
\newcommand{\ei}{\end{itemize}}
\newcommand{\ben}{\begin{enumerate}}
\newcommand{\een}{\end{enumerate}}
\newcommand{\bc}{\begin{center}}
\newcommand{\ec}{\end{center}}
\newcommand{\bt}{\begin{table}}
\newcommand{\et}{\end{table}}
\newcommand{\btb}{\begin{tabular}}
\newcommand{\etb}{\end{tabular}}
\newcommand{\bvec}{\left ( \ba{c}}
\newcommand{\evec}{\ea \right )}

\newcommand\eurA{\eur{A}}
\newcommand\scrA{\mathscr{A}}

\newcommand\eurB{\eur{B}}
\newcommand\scrB{\mathscr{B}}

\newcommand\eurV{\eur{V}}
\newcommand\scrV{\mathscr{V}}
\newcommand\scrW{\mathscr{W}}

\newcommand\eurD{\eur{D}}
\newcommand\eurJ{\eur{J}}
\newcommand\eurL{\eur{L}}
\newcommand\eurW{\eur{W}}

\newcommand\eubD{\eub{D}}
\newcommand\eubJ{\eub{J}}
\newcommand\eubL{\eub{L}}
\newcommand\eubW{\eub{W}}

\newcommand\bmupalpha{\bm\upalpha}
\newcommand\bmupbeta{\bm\upbeta}
\newcommand\bmuppsi{\bm\uppsi}
\newcommand\bmupphi{\bm\upphi}
\newcommand\bmuprho{\bm\uprho}
\newcommand\bmupxi{\bm\upxi}

\newcommand\calJ{\mathcal{J}}
\newcommand\calL{\mathcal{L}}

\newcommand{\notyet}[1]{{}}

\newcommand{\sgn}{\mathop{\rm sgn}}
\newcommand{\rk}{\mathop{\rm rk}}
\newcommand{\rank}{\mathop{\rm rank}}
\newcommand{\corank}{\mathop{\rm corank}}
\newcommand{\range}{\mathop{\rm Range\,}}
\newcommand{\supp}{\mathop{\rm supp}}
\newcommand{\p}{\partial}
\renewcommand{\P}{\grave{\partial}}
\newcommand{\yDelta}{\grave{\Delta}}
\newcommand{\yD}{\grave{D}}
\newcommand{\yeurD}{\grave{\eur{D}}}
\newcommand{\yeubD}{\grave{\eub{D}}}
\newcommand{\at}[1]{\vert\sb{\sb{#1}}}
\newcommand{\At}[1]{\biggr\vert\sb{\sb{#1}}}
\newcommand{\vect}[1]{{\bold #1}}
\def\R{\mathbb{R}}
\newcommand{\C}{\mathbb{C}}
\def\hvar{{\hbar}}
\newcommand{\N}{\mathbb{N}}\newcommand{\Z}{\mathbb{Z}}
\newcommand{\Abs}[1]{\left\vert#1\right\vert}
\newcommand{\abs}[1]{\vert #1 \vert}
\newcommand{\Norm}[1]{\left\Vert #1 \right\Vert}
\newcommand{\norm}[1]{\Vert #1 \Vert}
\newcommand{\Const}{{C{\hskip -1.5pt}onst}\,}
\newcommand{\sothat}{{\rm ;}\ }
\newcommand{\Range}{\mathop{\rm Range}}
\newcommand{\ftc}[1]{$\blacktriangleright\!\!\blacktriangleright$\footnote{AC: #1}}

% Italic ``theorems''
%\theoremstyle{plain}
\newtheorem{lemma}{Lemma}[section]
\newtheorem{hypothesis}[lemma]{Hypothesis}
\newtheorem{corollary}[lemma]{Corollary}
\newtheorem{proposition}[lemma]{Proposition}
\newtheorem{claim}[lemma]{Claim}

\newtheorem{theorem}[lemma]{Theorem}

% Roman ``theorems''
%\theoremstyle{definition}
\newtheorem{definition}[lemma]{Definition}
\newtheorem{assumption}[lemma]{Assumption}

% Humble things: remarks and examples.
%\theoremstyle{remark}
\newtheorem{remark}[lemma]{Remark}
\newtheorem{example}[lemma]{Example}
\newtheorem{problem}[lemma]{Problem}
\newtheorem{exercise}[lemma]{Exercise}

\newcommand{\const}{\mathop{\rm const}}

\renewcommand{\theequation}{\thesection.\arabic{equation}}

\makeatletter\@addtoreset{equation}{section}
%\makeatletter\@addtoreset{theorem}{section}
%\makeatletter\@addtoreset{proposition}{section}
%\makeatletter\@addtoreset{lemma}{section}
%\makeatletter\@addtoreset{corollary}{section}
%\makeatletter\@addtoreset{remark}{section}
%\makeatletter\@addtoreset{assumption}{section}
%\makeatletter\@addtoreset{definition}{section}
\makeatother

\def\Tau{\mathcal{T}}

\def\os{{o}}
\def\ol{{O}}
\def\dist{\mathop{\rm dist}\nolimits}
\def\spec{\sigma}
\def\mod{\mathop{\rm mod}\nolimits}
\renewcommand{\Re}{\mathop{\rm{R\hskip -1pt e}}\nolimits}
\renewcommand{\Im}{\mathop{\rm{I\hskip -1pt m}}\nolimits}

%\input{t1ptm.fd}
%\input{ulasy.fd}
%\input{ueus.fd}
%\input{ueuf.fd}

%\input{ursfs.fd}

%\title{\boldmath Emergent Horava gravity in graphene}

%% %simple case: 2 authors, same institution
% \author{G.E. Volovik$^{a,b}$,}
%% \author{and A. Nother Author}
% \affiliation{$^{a}${ Olli Lounasmaa Laboratory, School of Science and Technology,
%Aalto University, Finland} \\
%$^{b}${ L.D. Landau Institute for Theoretical Physics, Moscow, Russia}}

% more complex case: 4 authors, 3 institutions, 2 footnotes
%\author{M.A.Zubkov$^{c}$}

% The "\note" macro will give a warning: "Ignoring empty anchor..."
% you can safely ignore it.

%\affiliation{$^{c}$Institute for Theoretical and Experimental Physics, Moscow, Russia}

% e-mail addresses: one for each author, in the same order as the authors
%\emailAdd{zubkov@itep.ru}

%\begin{document}
%\maketitle
%\flushbottom

\def\w{\omega}
\def\o{\omega}

\newcommand{\bu}{{\bf \delta}}
\newcommand{\bk}{{\bf k}}
\newcommand{\bq}{{\bf q}}
\def\({\left(}
\def\){\right)}
\def\[{\left[}
\def\]{\right]}

\section{Introduction}

It can be shown \cite{Horava2005,NielsenNinomiya1981,Froggatt1991,Volovik2003,Volovik1986A,Volovik2011}, that in $3+1$ D systems with fermions near the Fermi - points the two - component spinors $\psi$ effectively appear. The remaining components of the original fermion field are in general case massive, and, therefore, decouple in the low energy limit. These Weil spinors are described by action
\begin{equation}
{\cal S} = \int d^{D} x |{\rm det}\,e| \, e^{\mu}_a \, \bar{\psi}
\sigma^a(p_\mu- {\cal A}_\mu)\psi +  \ldots \label{Hamiltonian}
\end{equation}
In this action  the collective mode
${\cal A}_\mu$ may contain the shift of the Fermi point resulted in the
effective gauge field $A_{\mu}$ and the effective spin connection
$\frac{i}{8} C^{ab}_{\mu} [\sigma^a,\sigma^b]$. The modes $e^{\mu}_a$ may be considered as the vielbein describing the gravitational degrees of freedom\footnote{It is worth mentioning that there is the conceptually different approach to emergent
gravity, in which the bilinear combinations  $\psi \hat O^{a}_i \psi $ or $\bar \psi \hat O^a_i \psi$ are identified with the vielbein for some tensorial operators $\hat O$. Corresponding constructions were considered in the context of relativistic field theory in \cite{Wetterich2004,Diakonov2011} and in the context of condensed matter physics in \cite{Volovik1986}.}.
Here we demonstrate, that the same phenomenon occurs in the $2+1$ D model of monolayer graphene.
The corresponding field ${\cal A}_\mu$ contains the $U(1)$ gauge field and does not contain any spin connection (in the leading approximation in elastic deformations). The corresponding mode $e^{\mu}_a$ contains the nontrivial $2D$ zweibein. Both these emergent fields are expressed through strain.

\section{The tight - binding model with varying hopping parameters. Floating Fermi - point as the emergent gauge field.}

\label{sectmono1}

%In \cite{Vozmediano,Vozmediano1} it was considered how the elastic deformations affect the tight - binding model of monolayer graphene.
 The carbon atoms of graphene form a
honeycomb
lattice with two sublattices A and B (of the triangular form). We
denote the lattice spacing by $a$. Let us introduce vectors that connect a
vertex of the sublattice A to its neighbors (that belong to the sublattice
B): ${\bf l}_1= (-a,0)$, ${\bf l}_2 =  (a/2,a\sqrt{3}/2)$, ${\bf l}_3=
(a/2,-a\sqrt{3}/2)$.

We suppose that the hopping parameter varies, so that its value
depends on the particular link connecting two adjacent points of the
honeycomb lattice. This is caused by elastic deformations. We have three values of $t_a, a = 1,2,3$ at each
point.
The Hamiltonian has the form
\begin{equation}
H=-\sum_{\alpha\in A}\sum_{j=1}^3 t_j({\bf r}_\alpha)
    \Bigl(\psi^\dag ({\bf r}_\alpha) \psi({\bf r}_\alpha + {\bf l}_j)
        + \psi^\dag ({\bf r}_\alpha +{\bf
l}_j)\psi({\bf r}_\alpha)\Bigr)\,,\label{H12}
\end{equation}

In addition, we
define the following variables:
${\bf m}_1= -{\bf l}_1 + {\bf l}_2, \qquad {\bf m}_3 = -{\bf
l}_3 + {\bf l}_1,  \qquad {\bf m}_2=
-{\bf l}_2 + {\bf l}_3 = -{\bf m}_1 - {\bf m}_3
$.
 In momentum space the effective  Hamiltonian has
has the form $H =  \int \frac{d^2k}{\Omega} \frac{d^2k^{\prime}}{\Omega} \psi^\dag({\bf k}^{\prime})\, \hat{U}\Bigl(t_j({\bf k}_-), {\bf k}_+, {\bf k}_-\Bigr) \, \psi({\bf
k})$, where
\begin{equation}
\hat{U}(t_j, {\bf k}_+, {\bf k}_-) = - \sum_{j=1}^3 t_j \left(\begin{array}{cc} 0 &   e^{-i{\bf
l}_j ({\bf k}_+ + {\bf k}_-)} \\
e^{i {\bf l}_j ({\bf k}_+ - {\bf k}_-)} & 0 \end{array}\right)\label{V1}
\end{equation}
while $\Omega$ is the area of momentum space, ${\bf k}_+ = ({\bf k} + {\bf k}^{\prime})/2$ and ${\bf k}_- = ({\bf k} - {\bf k}^{\prime})/2$.
We imply, that the variations of $t_a$ are given by
\begin{equation}
t_a({\bf r}) = t (1 - \Delta_a({\bf r})), \quad |\Delta_a| \ll 1
\end{equation}

 We  introduce the notation $K^{\pm}(t_j)$ for the "floating Fermi point". It is defined in such a way, that for ${\bf k}_+ = K^{\pm}(t_j)$, ${\bf k}_-=0$ the function $\hat{U}(t_j, {\bf k}_+, {\bf k}_-)$ vanishes: $\hat{U}(t_j,K^{\pm}(t_j), 0) = 0$.
For the case of homogenious variations of hopping parameters when ${\bf k}_- = 0$ this definition gives the true Fermi point.  For ${\bf k}_- \ne 0$ the interpretation is not so obvious. In general case the eigenvectors of the one - particle Hamiltonian do not correspond to the definite value of momentum. We have the complicated wave - packets instead. However, in the next section it will be shown that in the linear approximation  of $\hat{U}(t_j, {\bf k}_+, {\bf k}_-)$ (as a function of ${\bf k}_+$, ${\bf k}_-$) the term proportional to ${\bf k}_-$ may be neglected in the considered theory. This brings us back to the interpretation of $K^{\pm}(t_j)$ as the Fermi point. Now we take into account that hopping parameters $t_j$ themselves depend on ${\bf k}_-$. That's why $K^{\pm}(t_j({\bf k}_-))$ is the momentum - dependent ("floating") Fermi point. One can easily find $K^{\pm} \approx K^{(0)}_{\pm} \pm {\bf A}$,
where in addition to the fixed Fermi - point $K^{(0)}_{\pm} = \pm \frac{4\pi}{3\sqrt{3}}\Bigl(-\frac{{\bf m}_2}{\sqrt{3}}\Bigr)$ the emergent $U(1)$ gauge field ${\bf A}$ appears with the components \cite{sato}:
\begin{eqnarray}
{\bf A}^b &=&  - \frac{2}{3a^2} \epsilon^{ba} \sum_j  \Delta_{j} {\bf l}_j^a\label{AFP2}
\end{eqnarray}

\section{Expansion near the floating fermi - point}

Next, we expand $\hat{U}(t_j, {\bf k}_+, {\bf k}_-)$ around ${\bf k}_+ = K^{\pm}(t_j)$, ${\bf k}_-=0$. The result has the form:
\begin{eqnarray}
 \hat{U}(t_j, {\bf k}_+, {\bf k}_-)&=& -i \sigma^3\Bigl[(\mp \sigma^2 {\bf f}_2 + \sigma^1 {\bf f}_1  )\Bigl({\bf k}_+ - K^{\pm}(t_j)\Bigr)
  - ( \sigma^1 {\bf f}_1 \mp \sigma^2 {\bf f}_2  ) \sigma^3  {\bf k}_-\Bigr],\label{Hpm2Q}
\end{eqnarray}
where $\bf f$ (as well as $\bf A$) depends  on ${\bf k}-{\bf k}^{\prime}$ and is given by
\begin{eqnarray}
{\bf f}^k_{a}& = &    v_F \Bigl( \delta^k_a  - \frac{2}{3a^2} \sum_j  \Delta_{j} \Bigl[{\bf l}_j^a {\bf l}_j^k    - \frac{a}{2} {\bf l}^d_j \, K^{dak}\Bigr] \Bigr)\label{f}
\end{eqnarray}
Here we introduced \cite{Vozmediano} the new tensor $K$: $K^{ijk}=-\frac{4}{3a^3} \sum_b {\bf l}_b^i {\bf l}^j_b {\bf l}_b^k, \quad  K^{111}=-K^{122}=-K^{221}=-K^{212} = 1$.
One can see, that the emergent $U(1)$ field is related to the field $\bf f$ as follows:
\begin{eqnarray}
{\bf A}^i & = &  -\frac{1}{2v_Fa}\epsilon^{ik} K^{kjb} {\bf f}^j_b\label{a}
\end{eqnarray}

We define the new spinors: $\Psi_{\pm}({\bf Q}) = \psi(K^{(0)}_{\pm} + {\bf Q})$.
In order to return to the coordinate space we should use the following rule $\frac{1}{2} ({\bf Q} + {\bf Q}^{\prime}) F({\bf Q} - {\bf Q}^{\prime}) \rightarrow  -\frac{i}{2} ( F({\bf x}) \overrightarrow{\nabla} - \overleftarrow{\nabla} F({\bf x})),
\frac{1}{2} ({\bf Q} - {\bf Q}^{\prime}) F({\bf Q} - {\bf Q}^{\prime}) \rightarrow  - \frac{i}{2} \Bigl(\nabla F({\bf x})\Bigr)$. As a result the hamiltonian has the form $H =  \sum_{\pm}\int d^2 x [\Psi^{\pm}({\bf x})]^\dag {\bf H}_{\pm}
\Psi^{\pm}({\bf x})$,
where
\begin{eqnarray}
{\bf H}_- =  -\sigma^3\, {\bf f}_a^k({\bf x}) \sigma^a\circ
[\partial_k + i ({\bf A}_k({\bf x}) +  \tilde{\bf A}_k({\bf x}))]; \,
{\bf H}_+  =- \sigma^2 \Bigl( \sigma^3 \, {\bf f}_a^k({\bf x}) \sigma^a\circ
[\partial_k - i ({\bf A}_k({\bf x}) +  \tilde{\bf A}_k)]\Bigr) \sigma^2\nonumber
\end{eqnarray}
Here the field $\bf f$ is defined in coordinate space and is related to the variables $\Delta_a$ according to  Eq. (\ref{f}).
The product $i {\bf f}^k_a \circ \partial_k$ in these equations should be understood as ${\bf f}^k_a \circ i \partial_k = \frac{i}{2} \Bigl( {\bf f}^k_a  \overrightarrow{\partial_k} - \overleftarrow{\partial_k} {\bf f}^k_a \Bigr)$.
The additional gauge field $\tilde{\bf A}$ appears: $\tilde{\bf A}_a({\bf x}) =  \frac{1}{2v_F} \nabla_i{\bf f}^i_b({\bf x}) \epsilon_{ba}$.
This additional field $\tilde{\bf A}$ is to be compared with the emergent gauge field $\bf A$.
One can see that $\tilde{\bf A} \sim a \nabla {\bf A}$. Therefore, this is not reasonable to keep this additional field together with $\bf A$ in the field - theoretical description, where all dimensional quantities are to be much larger than the lattice spacing $a$. This shows that even in case of the variations of $t_a$ depending on the position in coordinate space we may omit the derivatives of $t_a(x)$ in the effective low energy field - theoretical Hamiltonian (that is we may omit the terms $\sim {\bf k}_-$ in momentum space representation).
The field ${\bf f}$ is related to the dreibein ${\bf e}$ as follows: $e \,{\bf e}_a^i = {\bf f}_a^i$; $e\, {\bf e}_0^0 = 1$, ${\bf e}_a^0 = {\bf e}_0^i = 0$,  where  $i,a=1,2$.
Here the determinant of the zweibein ${\rm det}\, {\bf e}_a^k = 1$, where $a,k = 1,2$. At the same time the three - dimensional determinant of ${\bf e}$ is equal to ${\rm det}\, {\bf e}^{(3\times3)} = {\bf e}^0_0 = 1/e$.  The three - volume element is $d^{(3)} V = d^2 {\bf r} d t e({\bf r},t)$. The value of $e$ is given by $e = [{\rm det}\, {\bf f}]^{1/2} = v_F(1- \frac{1}{3}(\Delta_2 + \Delta_3 + \Delta_1))$.

\section{Elastic deformations as a source of emergent gravity}
\label{sectelastic}

The graphene sheet is parametrized by variable $x_k, k = 1,2$. The classical elasticity theory has the displacements $u_a(x)$ as degrees of freedom ($a=1,2,3$).  The three - dimensional coordinates $y_a$ of the graphene sheet are given by
\begin{equation}
y_k(x) =  x_k + u_k(x), \quad k = 1,2; \quad y_3(x) = u_3(x)
\end{equation}

At $u_a=0$ the graphene is flat.
  The emergent metric of elasticity theory is given by $g_{ik} = \delta_{ik} + 2 u_{ik},~~u_{ik} = \frac{1}{2}\Bigl(\partial_i u_k +
\partial_k u_i +  \partial_i u_a \partial_k u_a\Bigr), \quad a = 1,2,3,
\quad i,k = 1,2$.
According to \cite{Vozmediano} the elastic deformations of graphene result in the change of the hopping elements, which determine the effective geometry experienced by fermions.
The simplest  connection between the deformations and the hopping elements $t_n$ ($n=1,2,3$), which is allowed by symmetry, is \begin{equation}
t_a({\bf r})=t[1 - \frac{\beta}{a^2} u_{ik}({\bf r})  {\bf l}^i_a {\bf l}^k_a] \,.
\label{HoppingElements}
\end{equation}

The dimensionless phenomenological parameter $\beta$ is determined by the microscopic physics. As it was mentioned above, the given consideration works for the displacements $u_a$ that are not necessarily small. However, we imply that $\beta |u_{ij}| \ll 1$. This is the requirement that the derivatives of $u_a$ are small being multiplied by $\beta$.

The emergent geometry and emergent $U(1)$ gauge field are given by $ {\bf e}^i_a  =  {\bf f}^i_a /\Bigl({\rm det}\, {\bf f} \Bigr)^{1/2}$ and Eq. (\ref{a}) with
$
 {\bf f}^i_a  =  v_F\left(\delta^i_a - \beta \left[\begin{array}{cc} u_{11} & u_{21}  \\
                                                            u_{12} &       u_{22}   \end{array}\right]\right)$.
This results in the usual expression for the strain - induced electromagnetic field (see \cite{Vozmediano,Vozmediano2010,Vozmediano0,VK2010,Vozmediano1} and references therein).  As for our values of ${\bf f}^k_a$, they differ from the expression for the anisotropic Fermi velocity calculated in \cite{Vozmediano}.
The zweibein is given by $ {\bf e}^i_a  =  {\bf f}^i_a /e$. It is constructed in such a way that the determinant of the zweibein ${\rm det}\, {\bf e}^{(2\times 2)} = 1$.
The $2+1$ D volume element $d^{(3)}V = e({\bf r},t) \, d^2 {\bf r} dt$  corresponds to the function $ {e} =  v_F (1 - \frac{\beta}{2} u_{aa})$.
This function is related to the $(^0_0)$ - component of the dreibein as ${\bf e}^0_0 = 1/e$. The determinant of the dreibein is equal to $1/e$.

%\end{enumerate}

\section{Conclusions}
\label{DiscussionSection}

 Here we considered the long - standing problem about the type of the geometry experienced by fermionic quasiparticles in graphene in the presence of elastic deformations. In some of the previous works on this subject it was supposed  that such an emergent geometry is  Riemannian \cite{Vozmediano0}. Later, the derivation of the space - dependent Fermi velocity in the presence of strain was undertaken \cite{Vozmediano}. Here we present the direct derivation of the emergent geometry.  In our approach the expansion of the effective hamiltonian is performed near to the "floating" Fermi point (where the hamiltonian vanishes in the limit of homogenious elastic deformations). We demonstrate, that the emergent geometry is described by the $2D$ zweibein. Also, the emergent $U(1)$ gauge field appears (its expression through strain coincides with the one derived previously \cite{sato}) while the spin connection does not appear in the approximation linear in elastic deformations\footnote{Since the spin connection is not forbidden by symmetry, it should appear in the next approximations.}. The emergent $U(1)$ field, in turn, is expressed through the zweibein. Thus we deal with the varying $2D$ Weitzenbock geometry expressed through the elastic deformation. This means that there is the emergent teleparallel gravity.

The authors kindly acknowledge discussions of emergent gravity with D.I.Diakonov, and useful correspondence with M.Vozmediano.
This work was partly supported by RFBR grant 11-02-01227, by the
Federal Special-Purpose Programme 'Human Capital' of the Russian Ministry of
Science and Education. GEV
acknowledges a financial support of the Academy of Finland and its COE
program,
and the EU  FP7 program ($\#$228464 Microkelvin).

\end{document}